\documentclass{aa}  

\usepackage{graphicx}
\usepackage{txfonts}
\usepackage{lipsum}
\usepackage{subcaption}         
\usepackage{xcolor}             
\usepackage{lscape}             
\usepackage{placeins}           
\usepackage{subcaption}
\usepackage{hyperref}

\begin{document}
\title{Evolution of fractality in centrally concentrated young clusters}
\titlerunning{Evolution of fractality in centrally concentrated young clusters}
\authorrunning{Akhmetali et al.}

\author{Almat Akhmetali\inst{1,2}\fnmsep\thanks{Corresponding author. Email: almat.akhmetali@nu.edu.kz}, Adilkhan Assilkhan\inst{1,3,4,5},  Mordecai-Mark Mac Low\inst{3,4}, Nurzhan Ussipov\inst{2,1}, Marat Zaidyn\inst{2}, Ernazar Abdikamalov\inst{1,6}, Alison Sills\inst{7}, Xiaoying Pang\inst{8,9} \and Bekdaulet Shukirgaliyev\inst{6,1,10}\fnmsep\thanks{Email: bekdaulet.shukirgaliyev@nu.edu.kz}}

\institute{
\inst{1} Energetic Cosmos Laboratory, Nazarbayev University, 53 Kabanbay Batyr Ave., Astana, 010000, Kazakhstan\\
\inst{2} Department of Electronics and Astrophysics, Al-Farabi Kazakh National University, 71 Al-Farabi Ave., Almaty, 050040, Kazakhstan\\
\inst{3} Department of Astrophysics, American Museum of Natural History, 200 Central Park W., New York, NY, 10024, USA\\
\inst{4} Department of Astronomy, Columbia University, 538 W 120th St., New York, NY, 10027, USA\\
\inst{5} Gumarbek Daukeyev Almaty University of Power Engineering and Telecommunications, 126/1 Baytursynuli St., Almaty, 050000, Kazakhstan\\
\inst{6} Physics Department, Nazarbayev University, 53 Kabanbay Batyr Ave., Astana, 010000, Kazakhstan\\
\inst{7}{Department of Physics and Astronomy,  McMaster University, 1280 Main Street West 
Hamilton, ON, L8S 4M1, Canada}\\
\inst{8}{Department of Physics, Xi’an Jiaotong-Liverpool University, 111 Ren’ai Road, Dushu Lake Science and Education Innovation
District, Suzhou 215123, Jiangsu Province, P.R. China}\\
\inst{9}{Shanghai Key Laboratory for Astrophysics, Shanghai Normal University, 100 Guilin Road, Shanghai 200234, PR China}\\
\inst{10} Heriot-Watt University Aktobe Campus, K. Zhubanov Aktobe Regional University, 263 Zhubanov Brothers St., Aktobe, 030000, Kazakhstan}

\date{Received September XX, 20XX}
 
  \abstract{

    We investigate the structural evolution of young star clusters forming within centrally concentrated molecular clouds. Our simulations use the \texttt{Torch} framework, which integrates the \texttt{FLASH} magnetohydrodynamics code with the \texttt{AMUSE} environment, enabling a self-consistent treatment of gas dynamics, star formation, stellar evolution, radiative transfer, and gravitational interactions. We quantify cluster structure using the $Q$ parameter for fractality and compute fractal dimensions via two methods: box-counting and correlation dimension. Our results show that clusters generally inherit fractal substructure from their parental clouds, which is typically erased within $\sim 2.5\,t_\mathrm{ff}$ through dynamical relaxation. Massive stars can induce the formation of secondary subclusters via feedback, with outcomes strongly dependent on stellar mass and formation timing. Interactions among subclusters, including mergers and dispersal, can extend fractal structure beyond $4\,t_\mathrm{ff}$. We also find systematic correlations between the fractality parameter $Q$ and the fractal dimension: fractality is positively correlated with both the correlation and box-counting dimensions, with the correlation dimension exhibiting a stronger correlation. These results demonstrate how stellar feedback and internal dynamics jointly shape the measurable fractal properties of embedded star clusters.

    }

   \maketitle
\nolinenumbers
\section{Introduction}

Most stars are born within giant molecular clouds, assembling in groups that can range from a few dozen to millions of members, known as star clusters \citep[see, e.g.,][]{lada2003embedded, portegies2010young, krause2020physics}. Current models suggest that such clusters emerge through the global hierarchical collapse of the clouds \citep{vazquez2017hierarchical, grudic2018top}. As a cloud undergoes gravoturbulent collapse~\citep{larson1981turbulence}, it fragments into dense, star-forming clumps, referred to as subclusters~\citep{mac2004control, mckee2007theory}. When gravitationally bound, these subclusters can merge over time and form a single, larger star cluster.

The dynamical evolution of star clusters is a complex process influenced by many factors. Both numerical simulations and observational studies suggest that clusters inherit a fractal spatial structure from their parental molecular clouds~\citep{cartwright2004statistical, clarke2010physics, sanchez2009spatial, andre2010filamentary, andre2014filamentary, kuhn2014spatial, jaehnig2015structural, arzoumanian2019characterizing, ballone2020evolution}. Over time, this initial fractality is gradually erased due to internal gravitational interactions and external influences. During this evolution, unbound clusters dissolve into the field, while bound clusters contract toward their centers, forming more radially concentrated distributions. However, this is only a general picture. The precise timescale over which clusters lose their initial fractal distribution and the factors that dominate this transformation are still not well understood.

Some young clusters (e.g., $\rho$ Ophiuchus) exhibit centrally concentrated morphologies as early as 1 Myr, suggesting rapid dynamical evolution in certain environments~\citep{cartwright2004statistical}. Conversely, substructured morphologies have been detected in much older clusters. For example,~\citet{sanchez2009spatial} reported significant substructure in NGC 1513 and NGC 1641, both older than 100 Myr. Recent N-body simulations by~\citet{daffern2020dynamical} further support the idea that, under specific conditions, clusters may develop a centrally condensed distribution within just a few million years. These contrasting findings highlight the diversity in cluster evolution pathways and underscore the need for more detailed studies of fractal structure of clusters.

Fractality has been extensively studied in observed systems and pure N-body simulations \citep{cartwright2004statistical,de2006fractal,maschberger2010properties,sun2022hierarchical, ussipov2024fractal,qin20253d,coenda2025global, akhmetali2026fractality}, yet only a limited number of works have addressed the fractal nature of star clusters forming in hydrodynamical simulations of collapsing turbulent molecular clouds. An early investigation was conducted by~\citet{schmeja2006evolving}, who applied the so-called $Q$ parameter, first introduced by~\citet{cartwright2004statistical}, to both observations of young embedded star clusters and smoothed-particle hydrodynamics (SPH) simulations from~\citet{schmeja2004protostellar}. They found that the $Q$ values obtained from simulations were comparable to those measured for real star clusters, and reported no significant correlation between the fractality of the sink particle distributions and the properties of the turbulent field imposed on the simulated collapsing molecular clouds. A similar study was presented by~\citet{maschberger2010properties}, who examined two simulations with cloud masses of $10^3$ and $10^4\,M_\odot$, performed by~\citet{bonnell2003hierarchical} and~\citet{bonnell2008gravitational}, respectively. In both cases, the resulting sink particle clusters initially formed with low $Q$ values (around 0.4 -- 0.5), indicative of a high degree of substructure. In the lower-mass, gravitationally bound cluster, $Q$ increased to values characteristic of a non-fractal distribution within a couple of free-fall times. In contrast, for the higher-mass, initially unbound cluster, $Q$ remained nearly constant. These results support the view that star clusters forming hierarchically begin with a fractal configuration, which is gradually erased through mergers and relaxation processes.

Low $Q$ values during the early stages of star formation were also reported by~\citet{girichidis2012importance}, although their study considered much smaller, highly unstable clouds ($\sim 100\,M_\odot$). They found a mild dependence of $Q$ on both the initial density profile and the mode of turbulence in the collapsing cloud. In contrast, hydrodynamical simulations by~\citet{parker2015spatial} and~\citet{gavagnin2017star} showed little effect of stellar feedback on $Q$. \citet{ballone2020evolution} simulated clusters with masses between $10^4$ and $10^5\,M_\odot$ and found all clusters remained fractal after two free-fall times. More recently,~\cite{Laverde-Villarreal_2025} reported that their clusters formed from clouds of $2 \times 10^4$, $8 \times 10^4$, and $3.2 \times 10^5\,M_\odot$ lose their fractal structure after approximately 2.5 free-fall times. Overall, these studies indicate that $Q$ generally evolves from low values, characteristic of a fractal structure, to higher values, reflecting a more uniform, non-fractal configuration.

In this paper, we explore the emergence of fractality in young clusters seeded within centrally concentrated molecular gas structures. Unlike previous models that often assume a pre-existing stellar distribution--primarily used to investigate the dynamical evolution of clusters following instantaneous gas expulsion \citep[e.g.,][]{shukirgaliyev2017impact, bek+2018, shukirgaliyev2019violent, 2019MNRAS.486.1045S,  Bek+2021, Kalambay2022,Kalambay2025, ussipov2024fractal, Abylay+2024,bissekenov2024exploring, Marina+2025, Weis+2025}--our work analyzes the stellar birth process within a dynamic environment. To achieve this, we utilize the suite of combined magnetohydrodynamic and $N$-body simulations from \citet{assilkhan2025centrally}, which captures the self-consistent formation of stars from a collapsing gas clump. This allows for a more realistic investigation of the early dynamical coupling between the stellar and gaseous components and its impact on the resulting fractal structure.

The rest of the paper is organized as follows: Section~\ref{Sec:Methods} describes the simulation dataset and the statistical methods used to quantify fractality. Section~\ref{Sec:Results} presents our analysis of the structural evolution, and Section~\ref{Sec:Conclusions} summarizes the main conclusions of the study.

\section{Methods}
\label{Sec:Methods}

\subsection{Simulations of star cluster formation}

We analyze the stellar distributions produced in the simulation suite of \citet{assilkhan2025centrally}, with the aim of quantifying the time evolution of their spatial structure and fractality. The simulations follow the formation and dynamical evolution of young stellar systems from initially gaseous, centrally concentrated molecular clouds.

The simulations were performed with the \texttt{Torch} framework \citep{wall2019collisional,wall2020modeling}, which couples the \texttt{FLASH} magnetohydrodynamics code \citep{fryxell2000flash,dubey2014evolution} to the \texttt{AMUSE} environment \citep{zwart2009multiphysics,pelupessy2013astrophysical,zwart2013multi,portegies_zwart_2019_3260650}. Stellar evolution was treated with \texttt{SeBa} \citep{portegies1996population,toonen2012supernova}, stellar dynamics was integrated with the fourth-order Hermite \(N\)-body code \texttt{ph4} \citep{mcmillan2012simulations}, and dynamically formed close systems were handled with \texttt{Multiples} and \texttt{SmallN} \citep{hut1995building,portegies2018astrophysical}. Further details of the numerical implementation are given in \citet{assilkhan2025centrally}.

The initial gas distribution follows the centrally concentrated cluster-formation prescription of \citet{shukirgaliyev2017impact}, which builds on the local star-formation model of \citet{parmentier2013local}. This prescription is used only to construct the initial gas density profile. The simulations are initialized without an embedded stellar population; stars form subsequently through sink-particle formation and IMF sampling during the hydrodynamical evolution. Therefore, the stellar structures analysed in this work are not imposed by the initial conditions, but emerge dynamically from the coupled gas, stellar-dynamical, stellar-evolution, and feedback processes.

\begin{figure}
\centering
\includegraphics[width=\columnwidth]{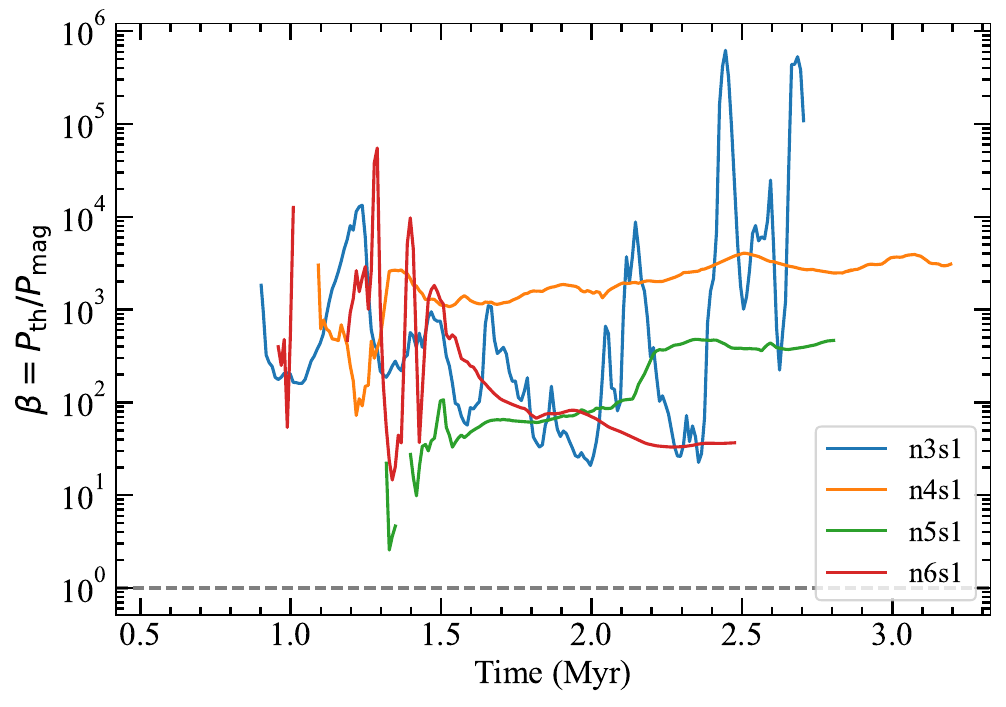}
\caption{Evolution of the $\beta=P_{\rm th}/P_{\rm mag}$, for the initial cloud with $B_z=3\times10^{-6}\,\mathrm{G}$. Values $\beta>1$ indicate that thermal
pressure exceeds magnetic pressure.}
\label{fig:beta_evolution}
\end{figure}

The initial cloud has a gas mass of \(M_{\rm gas}=2513\,M_\odot\), a spherical radius of \(R_{\rm sphere}=5.5\,\mathrm{pc}\), and a Plummer scale radius \(a_\star=1.1\,\mathrm{pc}\) used in the analytic density construction. The cloud is embedded in a uniform, initially static ambient medium, giving a total gas mass of \(3597\,M_\odot\) in the computational domain. The initial gas temperature is \(20\,\mathrm{K}\), the turbulent velocity field is normalized with a Kolmogorov spectrum \citep{wall2019collisional} to an initial gas virial parameter \(\alpha_{\rm vir,gas}=0.5\), and a weak vertical magnetic field of strength \(B_z=3\times10^{-6}\,\mathrm{G}\) is included. The corresponding magnetic pressure, $P_{\rm mag}=B^2/8\pi \simeq 3.6\times10^{-13}\,{\rm dyn\,cm^{-2}}$, whereas the ionized gas has $P_{\rm th}\sim10^{-11}\text{--}10^{-10}\,{\rm dyn\,cm^{-2}}$, yielding $\beta = P_{th}/P_{mag} \gg 1$; hence the magnetic field is dynamically negligible in the ionized regions where it can couple to the gas. Figure~\ref{fig:beta_evolution} illustrates the evolution of the $\beta$ during the early stages of the simulation.

We emphasize that the initial temperature \(T=20\,{\rm K}\) specifies the thermal internal energy of the cold molecular gas, not an effective temperature associated with turbulent kinetic motions. The turbulent kinetic energy is imposed independently through the initial turbulent velocity field, which is normalized to the adopted gas virial parameter. Thus, the thermal sound speed and the turbulent velocity dispersion are distinct components of the initial conditions.

The initial conditions for the gas set the initial mass-weighted temperature is \(\langle T\rangle_M\simeq19.8\,{\rm K}\), corresponding to an isothermal sound speed \(c_s=0.266\,{\rm km\,s^{-1}}\) for molecular gas. The gas velocity dispersion at the chosen gas virial parameter is larger than the thermal sound speed, with \(\sigma_{\rm gas,1D}\simeq0.77\,{\rm km\,s^{-1}}\) within \(R<5.5\,{\rm pc}\) and \(\simeq0.57\,{\rm km\,s^{-1}}\) within \(R<1.1\,{\rm pc}\). The corresponding one-dimensional Mach numbers are \(\mathcal{M}_{\rm 1D}\simeq2.9\) and \(\simeq2.1\), respectively. Therefore, the cloud is thermally cold but dynamically supersonic at initialization.

The simulation suite of \citet{assilkhan2025centrally}, consists of 13 models: ten independent realisations at refinement level \(n=3\), labelled \texttt{n3s1}--\texttt{n3s10}, and three higher-resolution models, \texttt{n4s1}, \texttt{n5s1}, and \texttt{n6s1}. The refinement levels \(n=3\)--6 correspond to maximum effective resolutions from \(64^3\) to \(512^3\) cells. The minimum cell size, sink accretion radius, and sink formation threshold were adjusted with resolution to resolve the local Jeans length. The simulation labels follow the convention \texttt{nXsY}, where \(X\) denotes the refinement level and \(Y\) denotes the random seed. The main initial parameters are summarized in Table~\ref{tab:Init1}.

\begin{table}[!ht]
\caption{Initial physical parameters of the simulation suite analysed in this work.}
\label{tab:Init1}
\centering
\small
\begin{tabular}{lc}
\hline\hline
Parameter & Value \\
\hline
Gas mass, \(M_{\rm gas}\) & \(2513\,M_\odot\) \\
Total gas mass in box & \(3597\,M_\odot\) \\
Sphere radius, \(R_{\rm sphere}\) & \(5.5\,\mathrm{pc}\) \\
Plummer radius used in analytic model, \(a_\star\) & \(1.1\,\mathrm{pc}\) \\
Minimum number of cells, Cells\(_{\rm min}\) & \(32^3\) \\
Maximum cell size, \(\Delta x_{\rm max}\) & \(0.429\,\mathrm{pc}\) \\
Temperature, \(T_{\rm cl}=T_{\rm amb}\) & \(20\,\mathrm{K}\) \\
Initial gas virial parameter, \(\alpha_{\rm vir,gas}\) & \(0.5\) \\
Vertical magnetic field, \(B_z\) & \(3\times10^{-6}\,\mathrm{G}\) \\
\hline
\end{tabular}
\end{table}

\subsection{Fractality}

To quantify the fractal structure of the simulated star clusters, we employ two commonly used metrics: the $Q$ parameter~\citep{cartwright2004statistical}, which is likely the most popular method to quantify fractality, and the fractal dimension $f_{\mathrm{dim}}$~\citep{imre2006}. The $Q$ parameter provides a measure of the cluster's spatial structure by comparing the normalized mean edge length of the minimum spanning tree to the mean separation between stars. Values of $Q < 0.8$ indicate a hierarchically substructured (fractal) distribution, while $Q > 0.8$ corresponds to a more centrally concentrated, radially smooth cluster~\citep{cartwright2004statistical,cartwright2009measuring,Allison2010, Hetem2019fractal}.

The fractal dimension $f_{\mathrm{dim}}$ quantifies the degree of clustering and substructure in the stellar distribution. Lower values of $f_{\mathrm{dim}}$ correspond to highly substructured, filamentary distributions, whereas higher values indicate a nearly uniform or centrally concentrated distribution~\citep{qin20253d}.  

For each simulated cluster, we compute $Q$ and $f_{\mathrm{dim}}$ at different evolutionary stages to investigate how the initial substructure evolves during the collapse and early dynamical evolution. These metrics allow us to track the dissolution of initial fractality and the emergence of a more relaxed, centrally concentrated configuration.  
 
\subsubsection{Q parameter}
\label{sec:Q}
The $Q$ parameter is commonly employed in the analysis of both observational data and simulations~\citep{schmeja2006evolving, bastian2009spatial, cartwright2009measuring, sanchez2009spatial, maschberger2010properties, parker2012characterizing, parker2014dynamics, parker2015spatial, ballone2020evolution, Laverde-Villarreal_2025, coenda2025global, akhmetali2026fractality}.
This parameter is defined using normalized quantities to standardize the calculation as 
\begin{equation}
Q = \bar{m}/\bar{s},
\end{equation}
where $\bar{m}$ is the normalized mean edge length of the minimum spanning tree, and $\bar{s}$ is the normalized mean interparticle separation.
The normalization for $s$ is the characteristic cluster size $r_{\mathrm{cl}}$, defined as the radius of a sphere in three dimensions (3D) or circle in two dimensions (2D) centered on the  center of mass of the cluster and enclosing 95\% of its  total mass. Following~\citet{cartwright2009measuring}, the minimum spanning tree mean edge length $m$ is normalized by $(\pi r_{\mathrm{cl}}^{2} / N_s)^{1/2}$ in 2D, and by $\left(\frac{4}{3} \pi r_{\mathrm{cl}}^{3} / N_s\right)^{1/3}$ in 3D, where $N_s$ is the total number of stars. We estimate the uncertainty of the Q parameter with the jackknife method~\cite{jackknife}.

Values of $Q$ close to zero indicate that the mean MST edge length, $\bar{m}$, is much smaller than the mean interparticle separation, $\bar{s}$. This corresponds to a highly subclustered distribution, where points form tight, isolated groups with small $\bar{m}$ separated by large distances $\bar{s}$. In contrast, higher $Q$ values represent a smooth, uniform distribution in which $\bar{m}$ and $\bar{s}$ are comparable. Generally, values of $Q$ below a threshold of 0.8 for 2D distributions, or $0.7$ for 3D, indicate substructured morphologies, whereas values of $Q$ above these thresholds correspond to smooth, centrally concentrated distributions. A $Q$ value close to the threshold suggests an approximately uniform density profile.

\subsubsection{Fractal dimension}

Fractal dimension quantitatively characterizes the spatial arrangement of stars within a cluster. This metric reflects the degree of structural irregularity: lower fractal dimension values correspond to more clumpy and substructured morphologies, while higher values indicate smoother, more centrally concentrated distributions~\citep{qin20253d}.  We calculate the fractal dimension using two different methods, box counting and neighbor counting.

\begin{figure*}[h!]
\centering
\includegraphics[width=0.8\textwidth]{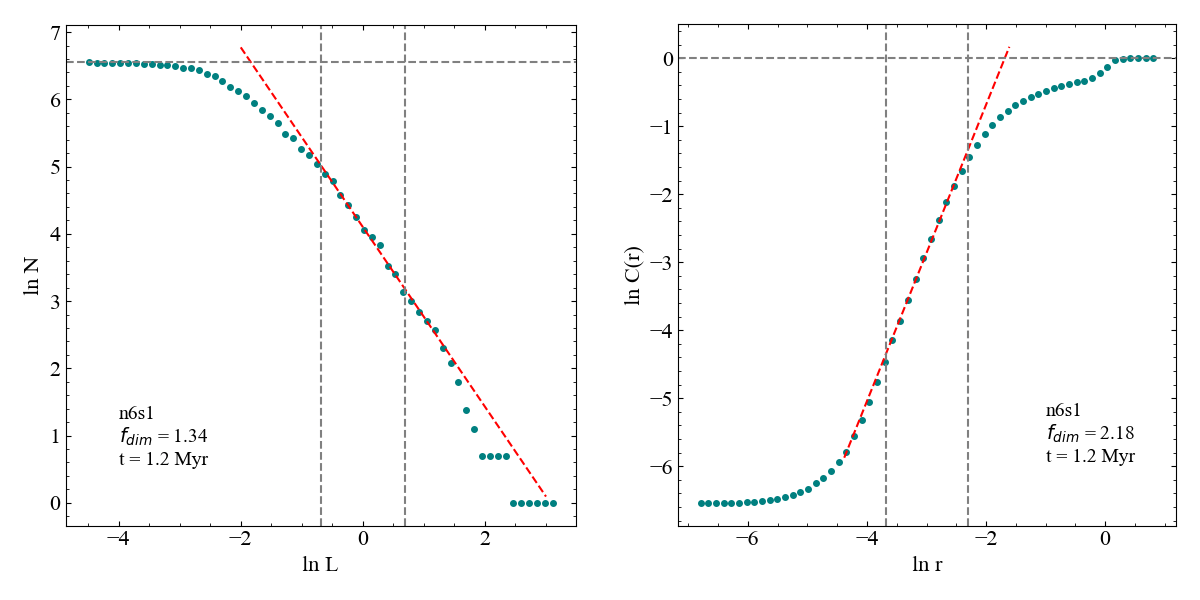}
  \caption{{\em Left:} Example of the derivation of the box-counting dimension from a plot of $\ln N(L)$ vs.\ $\ln L$ for the \texttt{n6s1} model in 3D. The {\em red solid line} represents the best-fit linear regression, with its slope corresponding to the estimated fractal dimension $f_\mathrm{dim}$. The vertical dashed lines mark the fitting range of $[-\ln 2, \ln 2]$. The horizontal {\em dashed line} indicates the natural logarithm of the total number of member stars in the cluster $N_s$. {\em Right}: Example of the derivation of the correlation dimension from a plot of $\ln C(r)$ vs.\ $\ln r $ for the same model. The {\em red solid line} represents the best-fit linear regression, with its slope corresponding to the estimated fractal dimension $f_\mathrm{dim}$. The vertical {\em dashed lines} mark the fitting range of $[-\ln 40, \ln 0.1]$. The horizontal {\em dashed line} indicates the natural logarithm of the normalized number of maximum neighbors, $\ln 1$.}
     \label{fig:FD}
\end{figure*}

The first method, box counting \citep{grassberger1983}, yields the Minkowski–Bouligand dimension~\citep{imre2006}. This method involves covering the dataset with boxes of various sizes and counting how many are required to fully enclose the data for each size. Before applying this approach, the 3D and 2D coordinates of the stars are standardized to have zero mean and unit variance. This standardization ensures that the coordinates become dimensionless, and consequently, the box size $L$ used in the calculation is also dimensionless. To reduce the influence of distant outliers, only stars located within a radius $r_{\mathrm{cl}}$, defined as the radius enclosing 95\% of the cluster's mass around the cluster's center of mass, are considered in the calculation. We estimate the uncertainty of the fractal dimension with the jackknife method for both methods~\cite{jackknife}.

The fractal dimension $f_{\mathrm{dim}}$ derived by the box-counting method is defined as
\begin{equation}
    f_{\mathrm{dim}} = -\frac{d \ln N(L)}{d \ln L},
\end{equation}
where $L$ is the box size, and $N(L)$ is the number of boxes required to cover the spatial distribution of member stars for a given $L$. Since we are working with a finite set of discrete points, this derivative must be approximated using finite differences. Moreover, the estimated value of $f_{\mathrm{dim}}$ may vary with box size $L$. In our analysis, we estimate $f_{\mathrm{dim}}$ as the slope obtained through linear regression of $\ln N(L)$ versus $-\ln L$ (see Fig.~\ref{fig:FD}(a)).

Figure~\ref{fig:FD}(a) shows that plateaus appear when the number of boxes $N$ approaches the total number of member stars. To reduce the bias introduced by these plateaus and to standardize the computation, we adopted a fixed fitting range of $[-\ln 2, \ln 2]$, marked by the vertical dashed lines in the figure. This interval corresponds to box sizes ranging from half to twice the characteristic scale set by the standard deviation of the coordinates, which approximately corresponds to the half-mass radius of the cluster.

The second method, neighbor-counting, computes the correlation integral~\citep{de2006fractal, sanchez2007nature}, approximated by the discrete sum
\begin{equation}
C(r) = \frac{1}{N_s (N_s - 1)} \sum_{i=1}^{N_s} n_i(r),
\end{equation}
where $n_i(r)$ is the number of neighbors of the $i$-th star contained within a 3D sphere or 2D circle of radius $r$. As in the box-counting method, the coordinates were standardized to zero mean and unit variance prior to the calculation.

The number of neighboring stars $n_i(r)$ within a sphere of radius $r$ centred on the $i$-th star is calculated as  
\begin{equation}
n_i(r) = \sum_{\substack{j=1 \\ j \neq i}}^{N_s} \Theta\left(r - \left| \mathbf{x}_i - \mathbf{x}_j \right|\right),
\end{equation}
where $\Theta(x)$ is the Heaviside step function, defined as $\Theta(x) = 0$ for $x < 0$ and $\Theta(x) = 1$ for $x \geq 0$ and $\left| \mathbf{x}_i - \mathbf{x}_j \right|$ is the distance between stars, which can be calculated in both 2D and 3D. In our analysis, $f_{\mathrm{dim}}$ is estimated as the slope obtained through linear regression of $\ln C(r)$ versus $\ln r$ (see Fig.~\ref{fig:FD}(b)).

Figure~\ref{fig:FD}(b) shows that a plateau develops when the average number of neighboring stars approaches the theoretical maximum, $N_s-1$. The fitting ranges were therefore chosen to isolate the approximately linear regime of the $\ln C(r)$--$\ln r$ relation prior to the onset of this saturation. Since the scale at which the plateau appears shifts to smaller $r$ as the total number of stars increases, the extent of the linear regime depends on $N_s$. This occurs because, even after coordinate normalization, the mean interparticle separation decreases with increasing $N_s$, causing the correlation integral to reach saturation at smaller radii. Consequently, different fitting intervals were adopted: $[-\ln 10, \ln 0.5]$ for the $n3s1$--$n4s1$ models, $[-\ln 20, \ln 0.2]$ for $n5s1$, and $[-\ln 40, \ln 0.1]$ for $n6s1$.

As illustrated in Fig.~\ref{fig:FD}, the two methods can yield substantially different values of \(f_{\rm dim}\) for the same snapshot. For the \texttt{n6s1} cluster, the box-counting method gives \(f_{\rm dim}=1.34\), while the correlation dimension yields \(f_{\rm dim}=2.18\). This difference reflects the distinct structural properties probed by the two estimators. The box-counting dimension measures how efficiently the stellar distribution fills space and is therefore sensitive to large voids and diffuse outer regions, often producing lower values. In contrast, the correlation dimension is based on pair separations and is more strongly influenced by dense local substructures, typically resulting in higher values in hierarchically clustered systems~\citep{de2006fractal}. The two methods therefore provide complementary information about the spatial structure of the cluster across different scales.

\section{Results}
\label{Sec:Results}

Young clusters are generally thought to inherit their fractal structure from their parental molecular clouds, acquiring this property through the collapse of local clumps. Over time, dynamical interactions gradually erase this fractality. However, key questions remain: what factors drive the change in a cluster’s fractality, and on what timescale is it lost?

In this work, the term cluster refers to the full stellar population formed within the simulation volume at a given snapshot, whether or not gravitationally bound, while subclusters denote local overdensities within this population, regardless of whether they are gravitationally bound to the cluster potential.

\subsection{Formation of subclusters}

\begin{figure*}[h!]
\centering
\includegraphics[width=0.8\textwidth]{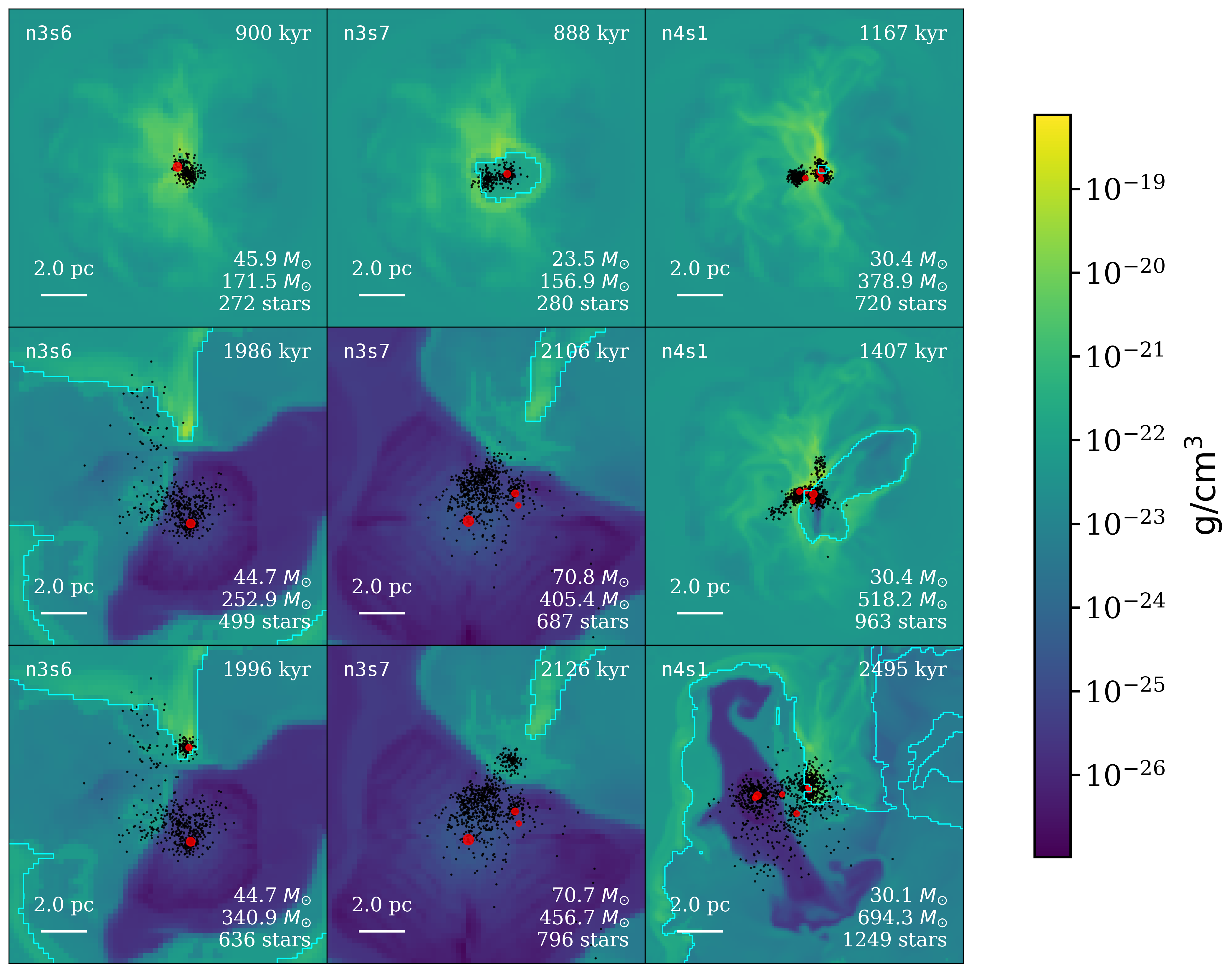}
  \caption{Gas density slices and projected stellar distributions for the \texttt{n3s6}, \texttt{n3s7}, and \texttt{n4s1} simulations, all of which exhibit subcluster formation. For \texttt{n3s6} and \texttt{n3s7}: {\em (top row)} onset of stellar feedback; {\em (middle row)} just before the formation of the second subcluster; {\em (bottom row)} after its formation. For \texttt{n4s1}: {\em (top row)} two subclusters are present; {\em (middle row)} they merge into a single system; {\em (bottom row)} the system separates again into two subclusters. Ionization fronts are marked by {\em cyan lines}. {\em Black dots} indicate individual stars with $M < 7\,M_\odot$, while massive stars ($M \geq 7\,M_\odot$) are shown in {\em red}, with symbol sizes proportional to stellar mass. Annotations in each panel show: {\em (bottom left)} scale bar; {\em (bottom right)} mass of the most massive star, total stellar mass within $5.5\,\mathrm{pc}$, and total number of stars; {\em (top left)} simulation label; and {\em (top right)} simulation time. The color scale indicates gas density in g\,cm$^{-3}$.
}
     \label{fig:Subclusters}
\end{figure*}

Models that form two or more distinct subclusters are presented in Figure~\ref{fig:Subclusters}, which shows the stellar distribution and gas density evolution for \texttt{n3s6}, \texttt{n3s7}, and \texttt{n4s1}. In \texttt{n3s6}, the first massive star is formed around 0.9 Myr, after which stellar feedback disperses the surrounding gas and displaces the dense region, allowing a second subcluster to form. 
In \texttt{n3s7}, two subclusters form around 0.9 Myr and start to merge into one, but then separate again around 2.1 Myr and form multiple subclusters. 
In \texttt{n4s1}, two subclusters form in close proximity at about 1.2 Myr, merge into a single system, and later separate again and form multiple subclusters.

These models show two scenarios of subcluster formation. First, as in  \texttt{n3s6}, a massive star forms in a single cluster and  expels the surrounding gas. The expelled gas sweeps up a dense shell where star formation occurs subsequently, forming the second subcluster. Second, as in the case of \texttt{n3s7} and \texttt{n4s1}, two subclusters form in close proximity, and start to merge until after dynamical relaxation they separate again and form a multiple subcluster system.

\subsection{Evolution of fractality}

\begin{figure*}[h!]
\centering
\includegraphics[width=1\textwidth]{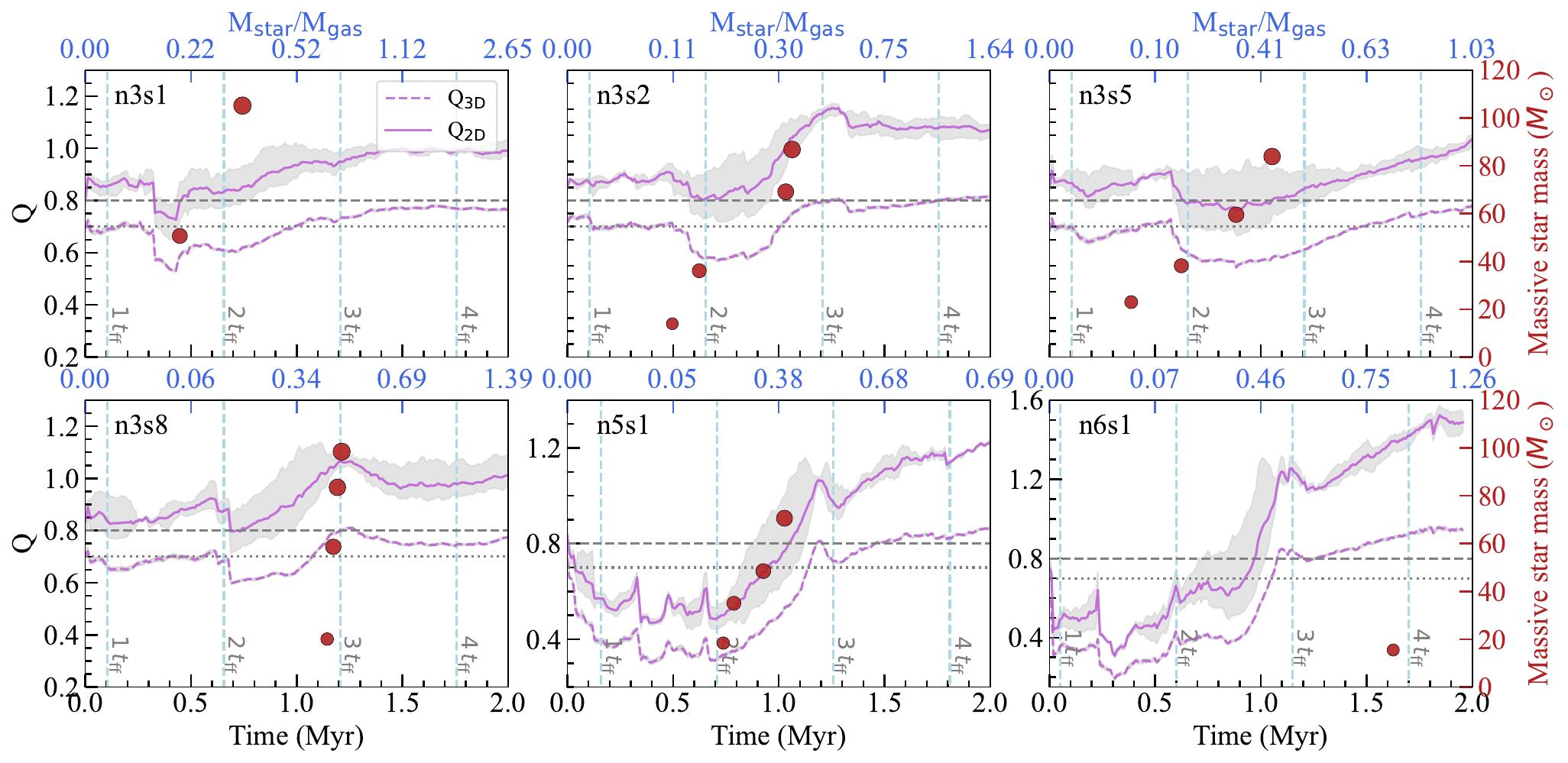}
    \caption{Evolution of the $Q$ parameter over time for models that initially exhibit fractality but subsequently evolve to smooth distributions. Time is measured starting from the onset of star formation. 
    {\em Dashed purple lines} show the $Q$ parameter calculated in 3D, with shaded areas indicating the uncertainty calculated with the jackknife method. (Note that in most cases the region of uncertainty is thinner than the plotted line.)  {\em Solid purple lines} represent the mean $Q$ values computed from 2D projections ($xy$, $yz$, and $xz$), with shaded areas indicating the minimum and maximum range of each projection.  
    {\em Red circles} represent the mass and time of formation of massive stars, with marker size and vertical position proportional to their mass; the secondary {\em red} $y$-axis on the right indicates the mass scale of the massive stars.  
    The top {\em blue} $x$-axis shows the instantaneous stellar-to-gas mass ratio at each time, $M_{\mathrm{star}}/M_{\mathrm{gas}}$.  
    {\em Vertical blue dashed lines} denote multiples of the cloud's free-fall time $t_\mathrm{ff}$, plotted with respect to the onset of star formation, while {\em horizontal gray lines} mark the fractality thresholds for the $Q$ parameter: 0.8 for 2D {\em (dashed line)} and 0.7 for 3D {\em (dotted line}) clusters.
     \label{fig:6Qs}}
\end{figure*}

Figure~\ref{fig:6Qs} shows the evolution of the fractality parameter $Q$ for models that show early fractal structure but eventually transition to more centrally concentrated states. For nearly all of these models, the critical point occurs at around $2.5\,t_\mathrm{ff}$, when both the 2D and 3D $Q$ values exceed their respective thresholds for fractal structure. Within our set of simulations, the transition from fractal to centrally concentrated configurations consistently occurs at $\sim 2.5\,t_\mathrm{ff}$, independent of numerical resolution and early fractal structure. Since all models share the same global density profile and mean volume density, this suggests that the dissolution of fractal substructure is primarily regulated by the global free-fall timescale of the parent cloud rather than by the initial degree of fractality.  A similar timescale was recently reported by \citet{Laverde-Villarreal_2025}, whose molecular clouds were an order of magnitude more massive than ours, suggesting that the loss of fractality may scale universally across different cloud masses. We note, however, that exploring different initial density profiles and mean densities will be required to assess whether this timescale truly remains invariant under more diverse initial conditions. Overall, these results indicate that fractal substructure typically dissolves on a timescale of a few free-fall times.

An important question is whether the formation of massive stars drives or modifies the dissolution of fractal substructure and the transition toward centrally concentrated stellar configurations. In some cases (e.g. \texttt{n3s2}, \texttt{n3s5}), the onset of massive star formation coincides with a sharp decrease in $Q$, consistent with feedback triggering secondary subclustering. However, this is not a universal outcome: in other models (\texttt{n3s1}, \texttt{n3s8}, \texttt{n5s1}) massive stars form without producing a measurable increase in fractality. Conversely, clusters such as \texttt{n6s1} lose their substructure over time even in the absence of massive stars. While feedback can enhance subclustering in some environments, the global evolution of $Q$ is governed by a combination of stellar feedback, dynamical relaxation, and the initial conditions of the parent cloud.

\begin{figure*}[h!]
\centering
\includegraphics[width=1\textwidth]{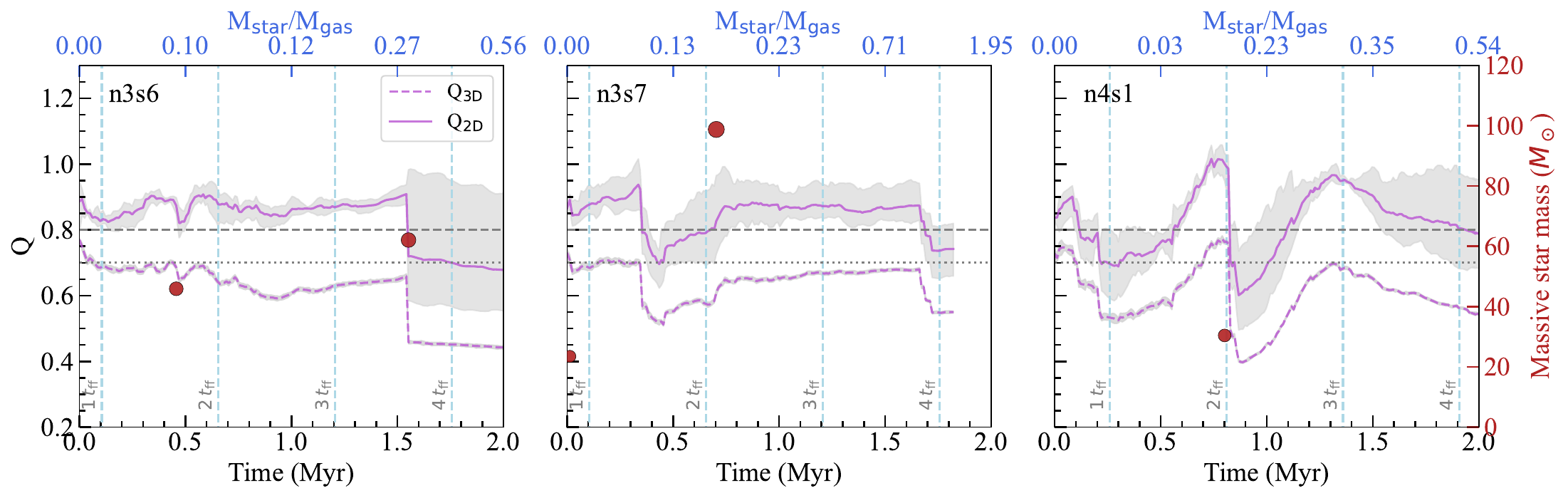}
    \caption{Evolution of the $Q$ parameter over time in models where subcluster formation maintained fractality after $4\,t_\mathrm{ff}$. All notation is the same as Figure~\ref{fig:6Qs}.
    \label{fig:3Qs}}
\end{figure*}

Figure~\ref{fig:3Qs} illustrates models that maintain fractality over longer timescales (snapshots of these models are shown in Figure~\ref{fig:Subclusters}). In these cases, $Q$ remains below the fractality threshold even after $4\,t_\mathrm{ff}$, indicating the long-term survival of subclustering. Rather than converging toward a smooth, centrally concentrated structure, these systems preserve a dynamically evolving, hierarchical morphology. This behaviour is typically associated with the repeated formation, merging, and reorganization of subclusters driven by gravitational interactions and stellar feedback.

In particular, the fluctuations in $Q$ observed for the \texttt{n4s1} model are caused by the transient merging and subsequent separation of two nearby subclusters. When the subclusters temporarily merge into a single system, the spatial distribution becomes more centrally concentrated and $Q$ increases. As the system later re-fragments into multiple subclusters, the degree of substructure increases again, leading to a decrease in $Q$. These oscillations demonstrate that $Q$ is sensitive not only to the presence of subclusters, but also to their dynamical interactions, making it an effective statistical tracer of time-dependent subclustering.

The initial spatial structure of the forming cluster is ultimately inherited from the turbulent density field of the parent molecular cloud, which forgets its initial Kolmogorov driving spectrum because it evolves in a crossing time to the $k^{-2}$ power spectrum set by supersonic turbulence (e.g. \citealt{mac2004control}), where $k$ is the wavenumber. The fragmentation of the gas into dense filaments and cores---driven by the turbulent spectrum---naturally imprints a hierarchical spatial distribution of star-forming sites, which is subsequently reflected in the initial configuration of sink particles. As a result, low values of the $Q$ parameter at early times are expected, as they directly trace the underlying fractal structure of the gas rather than stellar dynamical processes. During cluster evolution, this initial imprint is progressively modified by gravitational interactions, accretion, and feedback processes, but the early-time substructure remains strongly connected to the original turbulent power spectrum of the cloud.

\begin{figure}[h!]
\centering
\includegraphics[width=0.8\linewidth]{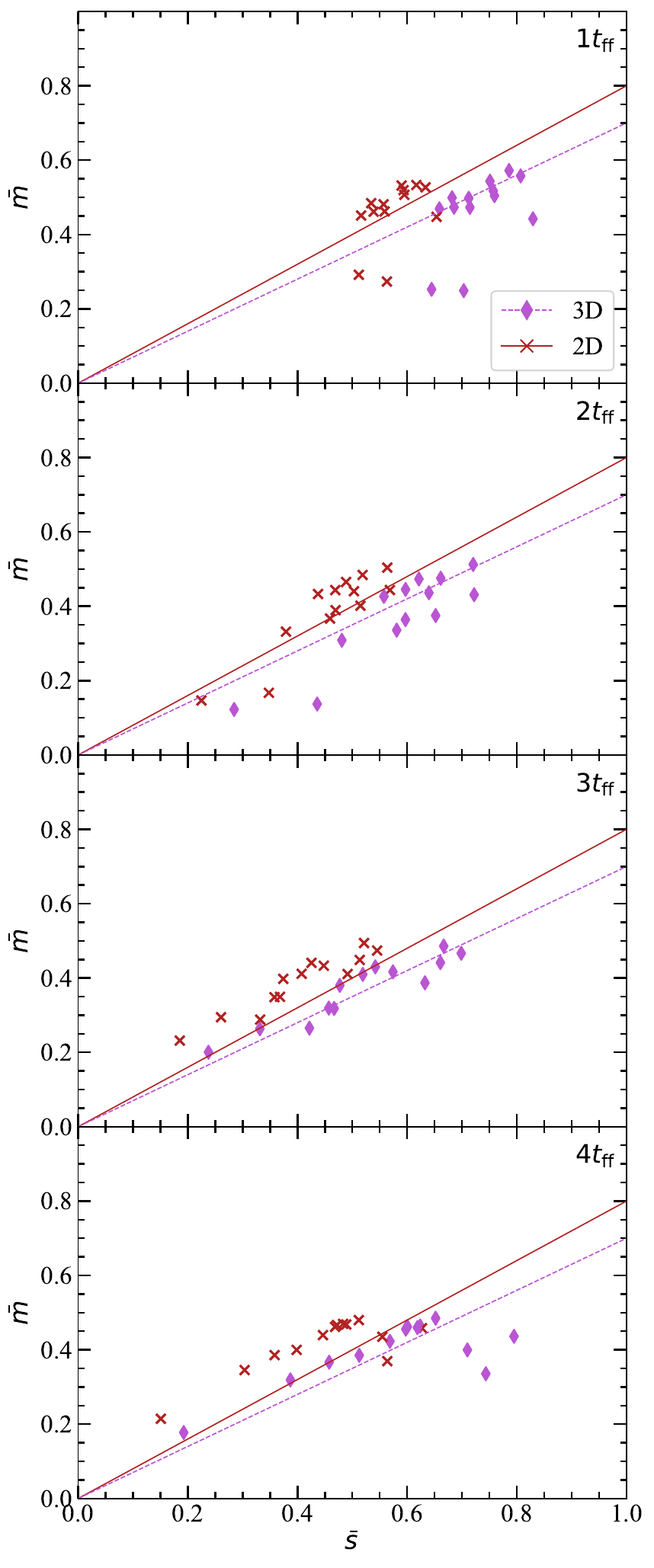}
    \caption{Mean interparticle distance $\bar{s}$ versus mean minimum spanning tree edge length $\bar{m}$ (see Section \ref{sec:Q}) for all simulations at different multiples of the free-fall time $t_\mathrm{ff}$, indicated in the upper right corner.  Fractality is indicated by points lying below and to the right of the critical lines.
    {\em Purple diamonds} and {\em dashed lines} represent 3D results, while {\em brown cross markers} with {\em solid lines} represent mean values of 2D projections ($xy$, $yz$, and $xz$). }
     \label{fig:s vs m}
\end{figure}

As demonstrated in \cite{cartwright2009measuring}, $\bar{s}$--$\bar{m}$ plots provide an even more sensitive diagnostic of fractality. Figure~\ref{fig:s vs m} can be directly compared with Figures 1 and 2 in ~\cite{cartwright2009measuring}. Values indicate fractality if they lie below the dashed line corresponding to their dimensionality. At early times ($1\,t_\mathrm{ff}$ and $2\,t_\mathrm{ff}$), most 3D values of $\bar{s}$ and $\bar{m}$ fall within the region corresponding to fractal distributions with $f_{\mathrm{dim}} \sim 2$ and $1.6$. However, for the 2D projections, most clusters lie in ranges consistent with radial density profiles following $n \propto r^\alpha$ with slopes $\alpha \sim -2.5$ and $-2$. By $3\,t_\mathrm{ff}$ and $4\,t_\mathrm{ff}$, both $\bar{s}$ and $\bar{m}$ decrease as the clusters evolve dynamically toward more centrally concentrated radial distributions. Consequently, the values of $\bar{s}$ and $\bar{m}$ converge toward those expected for radial density profiles with $\alpha \sim -2.9$ in both 2D and 3D cases, though three clusters continue to show clear fractal structure (n3s6, n3s7, and n4s1). 

We further compare the evolution of the $Q$ parameter obtained from our \texttt{n5s1} and \texttt{n6s1} models with the $N$-body simulations of \citet{Parker2017Nbody} and with the observational sample of young Galactic clusters from \citet{Dib2018Observation}. The $N$-body models consist of initially substructured stellar clusters with $N_\ast = 425$ and $N_\ast = 1500$, drawn from fractal spatial distributions with $D = 1.6$ and evolved from sub-virial initial conditions ($\alpha_{\rm vir} = 0.3$), covering initial local volume densities from $10$ to $10^{4}\,M_\odot\,\mathrm{pc}^{-3}$. The observational sample comprises young Milky Way clusters with $Q \approx 0.7$--$0.8$, corresponding to systems in early to intermediate dynamical stages.

For context, our \texttt{n5s1} and \texttt{n6s1} clusters reach mean stellar densities within $r_{\rm cl}$ of $\rho_\star \sim 10$ -$10^{3}\,M_\odot\,\mathrm{pc}^{-3}$ over the first $\sim 2\,t_\mathrm{ff}$, with central densities (within the half-mass radius) growing from $\sim 10^{2}$ to $\sim 10^{4}\,M_\odot\,\mathrm{pc}^{-3}$ as the clusters contract. The local stellar density measured around each star using the ten nearest neighbours \citep[Casertano-Hut estimator;][]{Casertano1985} has a median value $\tilde\rho_{10} \sim 10^{3}$--$10^{4}\,M_\odot\, \mathrm{pc}^{-3}$. These values place our models in the intermediate- to high-density regime of \citet{Parker2017Nbody}. We note, however, that central stellar densities of $\gtrsim10^{4}\,M_\odot\,\mathrm{pc}^{-3}$ are not uncommon among observed young clusters, and even denser systems are known~\citep{Larsen2004, Paumard2006}.

\begin{figure*}[h!]
\centering
\includegraphics[width=1\textwidth]{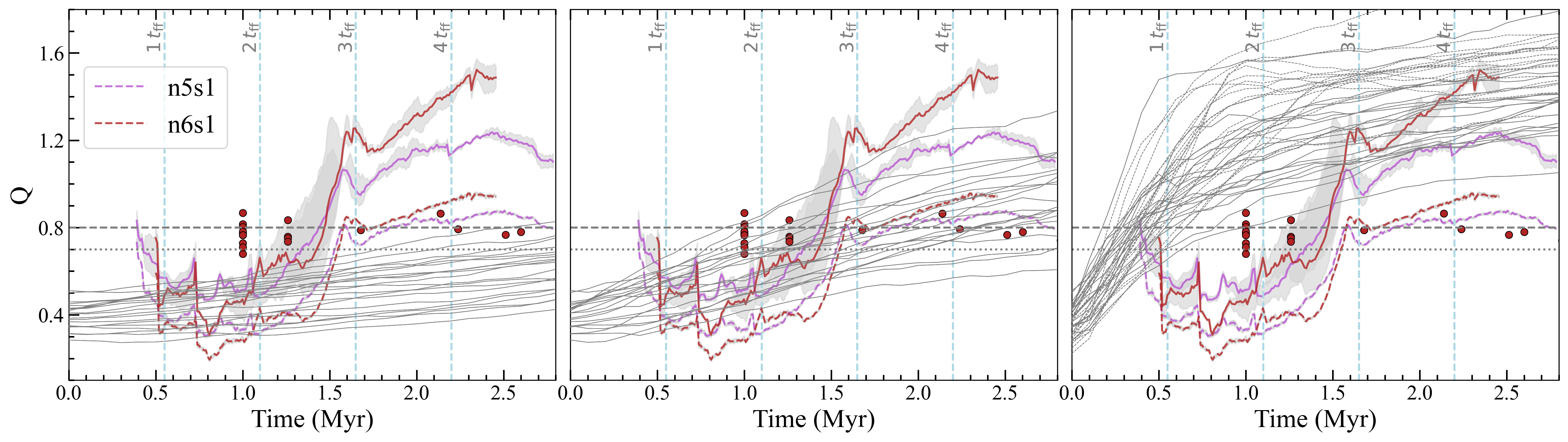}
    \caption{Comparison with $N$-body simulations and observational data of the $Q$ parameter over time for the {\tt n6s1} and {\tt n5s1} models.   
    {\em Dashed lines} show the $Q$ parameter calculated in 3D, with the shaded areas indicating the uncertainty calculated with the jackknife method. 
    {\em Solid lines} represent the mean $Q$ values computed from 2D projections ($xy$, $yz$, and $xz$), with shaded areas indicating the minimum and maximum range of the projections.   
    {\em Vertical blue dashed lines} denote multiples of the cloud's free-fall time $t_\mathrm{ff}$, while {\em horizontal gray lines} mark the fractality thresholds for the $Q$ parameter: 0.8 for 2D {\em (dashed line)} and 0.7 for 3D {\em (dotted line}) clusters.
    The $N$-body models include clusters with $N_\ast = 425$, in which the initial local volume densities lie in the range $10$--$60~M_\odot\,\mathrm{pc}^{-3}$ ({\em solid gray lines}, left panel), $100$--$500~M_\odot\,\mathrm{pc}^{-3}$ ({\em solid gray lines}, middle panel), and $5 \times 10^{3}$--$10^{4}~M_\odot\,\mathrm{pc}^{-3}$ ({\em solid gray lines}, right panel). Additionally, models with $N_\ast = 1500$ are included, with initial local volume densities in the range $2 \times 10^{3}$--$2 \times 10^{4}~M_\odot\,\mathrm{pc}^{-3}$ ({\em dashed gray lines}, right panel). The $Q$ parameters for observed clusters are given by {\em red circles}.The ages of the youngest observed clusters ($\sim1$ Myr) are subject to systematic uncertainties, and therefore the agreement between the observations and models at early times may be better than suggested by the nominal age estimates.}
    \label{fig:comparison_Q}
\end{figure*}

Figure~\ref{fig:comparison_Q} shows that, despite the very different numerical approaches, our models broadly reproduce the overall evolutionary trend seen in both the $N$-body simulations and the observations: $Q$ increases with time as the clusters evolve toward more centrally concentrated configurations. However, the $N$-body curves increase smoothly and quasi-monotonically, whereas our models display pronounced fluctuations throughout the evolution. We emphasize that the $Q$ parameter is a purely geometric descriptor of the stellar positions, and is therefore sensitive to any physical process that perturbs the spatial distribution of stars. In the $N$-body simulations of \citet{Parker2017Nbody}, the stellar positions evolve solely under collisionless gravitational dynamics, without stellar evolution, gas, or feedback. In contrast, in our simulations the stellar positions are shaped self-consistently by three coupled effects: the turbulent velocity field of the parent cloud sets the spatial distribution of newly-formed stars; the gravitational potential of the gas, which deepens and becomes more centrally concentrated as the cloud collapses, drags the embedded stars toward the centre; and stellar feedback from massive stars locally disrupts both the gas and the stellar component. The fluctuations of $Q$ in our models therefore reflect genuine hydrodynamical and feedback-driven rearrangements of the stellar component---such as the transient merging and re-separation of subclusters, and ionization-driven displacement of dense gas and embedded stars---rather than purely gravitational relaxation. Despite these qualitative differences in the underlying physics, the long-term trend toward centrally concentrated configurations is recovered in both approaches, with our \texttt{n5s1} and \texttt{n6s1} models showing the closest agreement with the $N_\ast = 425$ intermediate-density branch of \citet{Parker2017Nbody}, consistent with the stellar densities reached in our simulations.

\subsection{Evolution of fractal dimension}

\begin{figure*}[h!]
\centering
\includegraphics[width=1\textwidth]{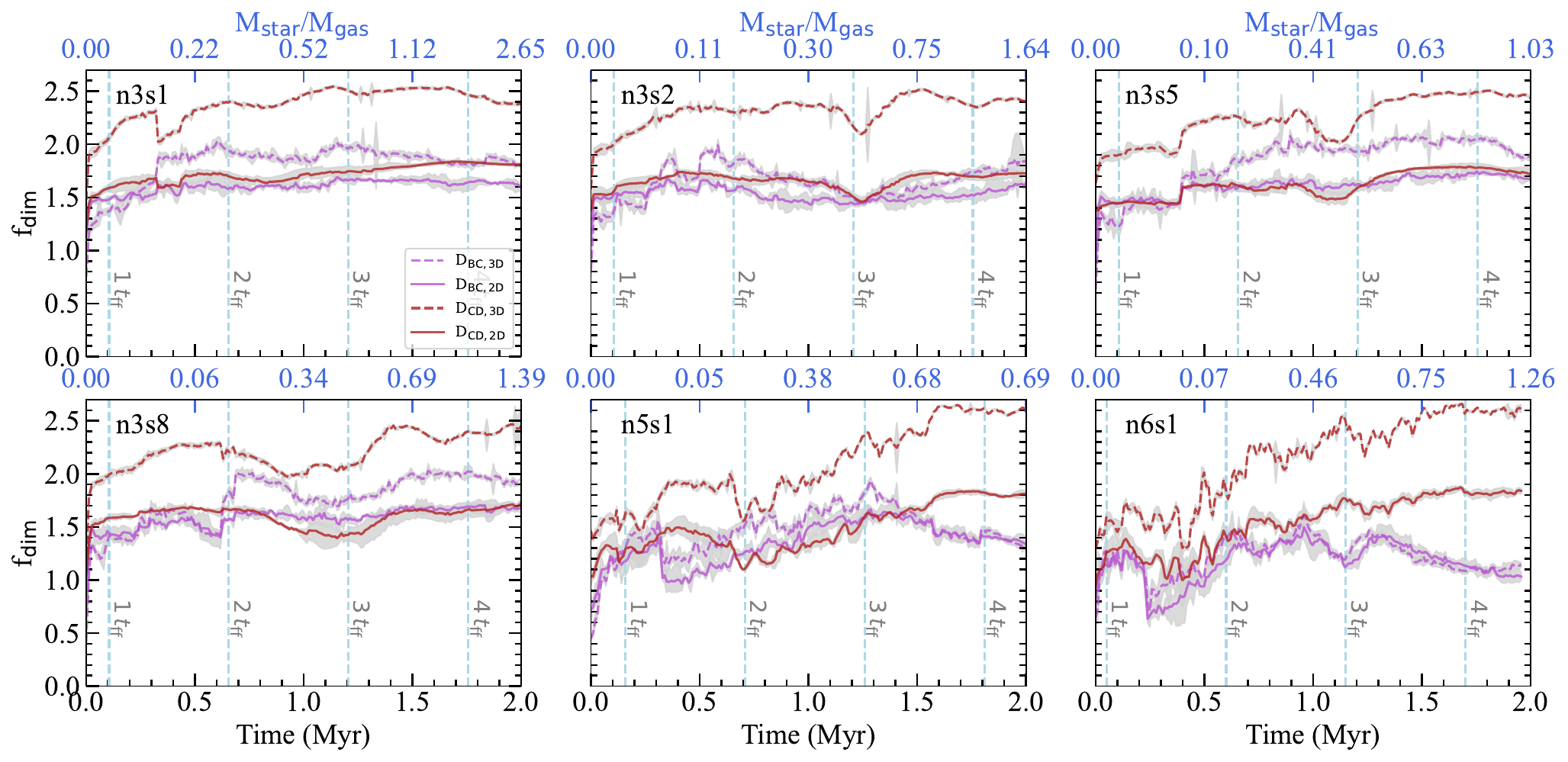}
    \caption{Evolution of the fractal dimension $f_{\mathrm{dim}}$ over time for the models that initially exhibit fractality but subsequently lose it shown in Figure~\ref{fig:6Qs}. Both the box counting dimension $D_\mathrm{BC}$ {\em (purple lines)}and correlation dimension $D_\mathrm{CD}$ {\em (brick red lines)} are shown. Time is measured starting from the onset of star formation.
    The secondary {\em blue} $x$-axis shows the stellar-to-gas mass ratio, $M_{\mathrm{star}}/M_{\mathrm{gas}}$. 
    {\em Dashed lines} show $f_{\mathrm{dim}}$ calculated in 3D, with the shaded regions indicating the uncertainty calculated with the jackknife method. {\em Solid lines} represent the mean $f_{\mathrm{dim}}$ values computed from 2D projections ($xy$, $yz$, and $xz$), with shaded areas indicating the minimum and maximum range of each projection. 
    Vertical {\em blue dashed lines} denote multiples of the cloud's free-fall time, plotted with respect to the onset of star formation}.  
    \label{fig:6Ds}
   
\end{figure*}

Figure~\ref{fig:6Ds} shows the evolution of $f_{\mathrm{dim}}$ for the models that initially exhibit fractal structure as evaluated by low Q values but later lose it, as shown in Figure~\ref{fig:6Qs}. In all models, $f_{\mathrm{dim}}$ increases sharply before $t_{\mathrm{ff}}$, as expected, since star formation proceeds rapidly while the clusters are still assembling and gaining structural complexity. Two distinct behaviors can be identified.
First, the \texttt{n3s*} models all show similar evolution. In these models, the correlation dimension in 3D $D_\mathrm{CD,3D}$ rises and reaches a plateau at approximately $2.4$, while the box-counting dimension $D_\mathrm{BC,3D}$ stabilizes at about $1.8\text{--}1.9$. In the 2D calculations, however, both methods give very similar results, with 
   both $D_\mathrm{CD,2D}$ and $D_\mathrm{BC,2D}$
converging to values around $1.5\text{--}1.6$.
Second, the higher-resolution models \texttt{n5s1} and \texttt{n6s1} show somewhat different behavior, particularly after $3\,t_{\mathrm{ff}}$. In these models, the box-counting dimension evolves similarly to the \texttt{n3s*} models but reaches a slightly lower plateau at about $1.3\text{--}1.4$. In contrast, the correlation dimension continues to increase and exceeds $2.5$ in 3D, while the corresponding 2D values remain within the same range as the other models.

The different behavior of the high-resolution models \texttt{n5s1} and \texttt{n6s1} is consistent with their large $Q$-parameter values ($Q \approx 1.2$), which indicate strongly centrally concentrated clusters with a radial density profile rather than a fractal structure. In such configurations, stars become densely packed toward the cluster center, significantly increasing the number of stellar pairs at small separations. Since the correlation dimension is based on pair statistics, this leads to a steeper correlation sum and consequently larger derived values of $f_{\mathrm{dim}}$ in 3D. In contrast, the box-counting dimension primarily reflects the global spatial extent of the cluster and is therefore less sensitive to strong central concentration, resulting in similar plateau values to those of the lower-resolution models. In projection, these differences are partially suppressed, which explains why the corresponding 2D estimates remain comparable across all simulations.
 
\begin{figure*}[h!]
\centering
\includegraphics[width=1\textwidth]{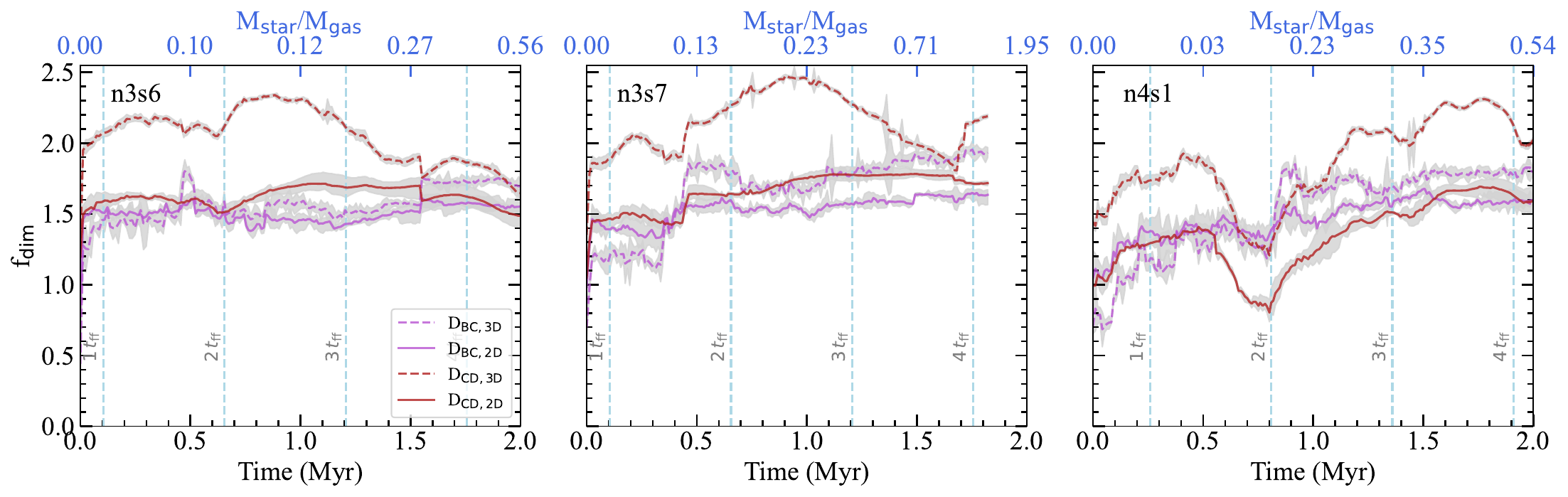}
    \caption{Evolution of the fractal dimension $f_{\mathrm{dim}}$ over time in models where the formation of subclusters allowed fractality as evaluated by low $Q$ values to be maintained even after $4\,t_\mathrm{ff}$ (see Fig.~\ref{fig:3Qs}).  All notations are the same as in Figure~\ref{fig:6Ds}.
    \label{fig:3Ds}}
\end{figure*}

Fig.~\ref{fig:3Ds} presents the evolution of $f_{\mathrm{dim}}$ over time for models that maintain their fractality as evaluated by low $Q$ values throughout the simulation (see Fig.~\ref{fig:3Qs}). In these models, $f_{\mathrm{dim}}$ increases during the first free-fall time and, after some fluctuations, plateaus at approximately $1.4$--$1.5$, similar to Figure~\ref{fig:6Ds}. However, a clear correlation is observed between sharp drops in $Q$ (Fig.~\ref{fig:3Qs}) and increases in $f_{\mathrm{dim}}$ around $1.5\,t_\mathrm{ff}$ and $2\,t_\mathrm{ff}$ for \texttt{n3s7} and \texttt{n4s1}, respectively. These sharp decreases in $Q$ and corresponding increases in $f_{\mathrm{dim}}$ coincide with the formation of the second subclusters in these models (Fig.~\ref{fig:Subclusters}, middle and right columns).

\subsection{Correlation of fractality and fractal dimension}

The $Q$ parameter indicates whether a system is fractal, whereas $f_{\mathrm{dim}}$ quantifies the degree of fractality. However, there is no widely accepted or consistent correlation between them, especially when comparing $Q$ with $f_{\mathrm{dim}}$ estimated using different methods. In this section, we examine the possible correlations between $Q$ and $f_{\mathrm{dim}}$ obtained from the box-counting method and the correlation sum.

\begin{figure*}[h!]
\centering
\includegraphics[width=0.95\textwidth]{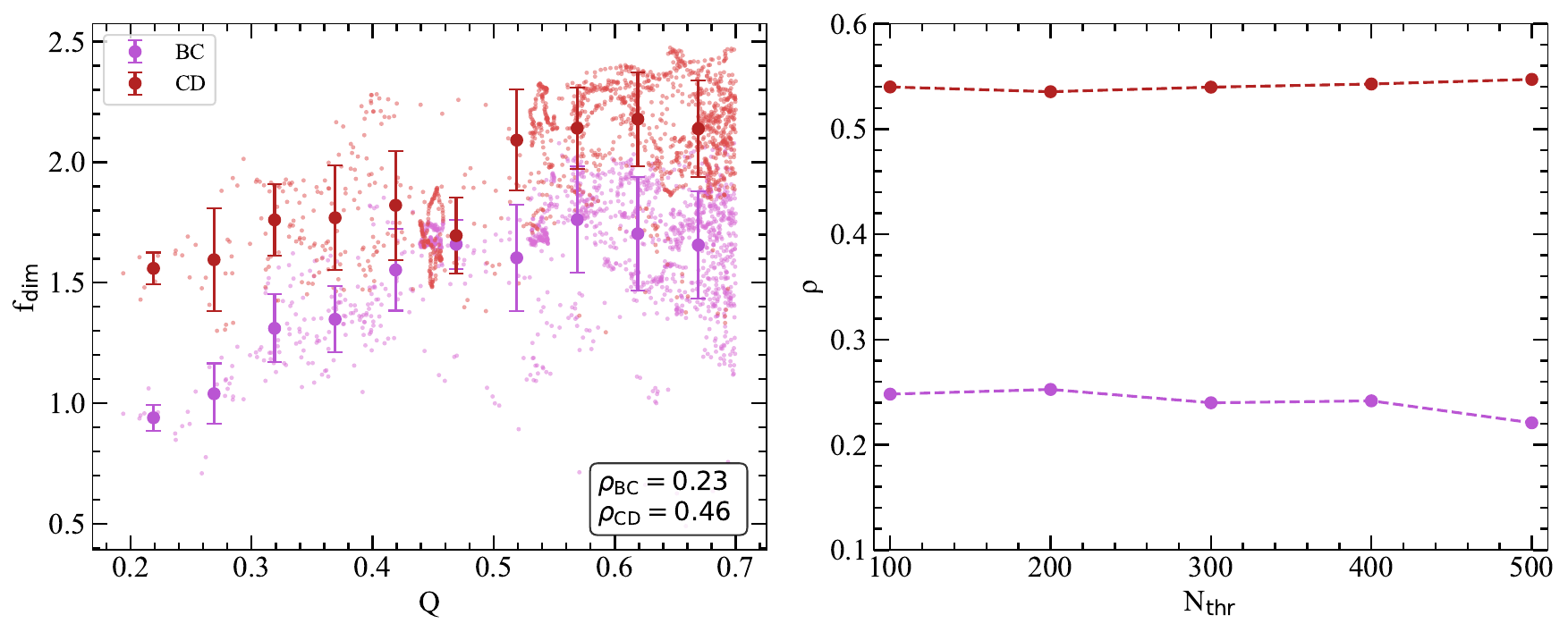}
    \caption{Correlation between the fractal dimension $f_{\mathrm{dim}}$ and the $Q$ parameter for the 3D stellar distributions. 
    {\em Left panel:} binned mean values of $f_{\mathrm{dim}}$ as a function of $Q$ for snapshots with $Q < 0.7$. 
    The box-counting dimension is shown in {\em purple} and the correlation dimension in {\em red}, with error bars indicating the standard deviation within each bin. Spearman’s rank correlation coefficient $\rho$ is shown for each dimension in the legend.
    Right panel: variation of $\rho$ as a function of the threshold number of stars per snapshot ($N_{\mathrm{thr}}$). }
    \label{fig:correlation}
\end{figure*}

Figure~\ref{fig:correlation} shows the dependence of $f_{\mathrm{dim}}$ on $Q$ parameter. We consider only snapshots with $Q < 0.7$, corresponding to fractal distributions. In the left panel, the points represent mean values with error bars showing the standard deviation. The correlation between the two variables was quantified with Spearman’s rank correlation coefficient $\rho$~\citep{zwillinger1999}.
By definition, $\rho=+1$ corresponds to a perfect positive correlation, $\rho=0$ to no correlation, and $\rho=-1$ to a perfect negative correlation. We find that $Q$ shows a weak positive correlation with the box-counting dimension and a moderate positive correlation with the correlation dimension. This is in tension with the findings of~\citet{sanchez2009spatial}, who found strong correlation between fractality and fractal dimension when applied to artificial clusters. 

The accuracy of $f_{\mathrm{dim}}$ estimation depends critically on the number of points considered, with larger $N_{thr}$ reducing uncertainties. We limit our analysis to snapshots with $N_{thr} \leq 500$, since for larger $N_{thr}$ the number of available snapshots decreases, preventing statistically robust conclusions.  The right panel of Figure~\ref{fig:correlation} shows the Spearman rank correlation coefficient $\rho$ as a function of the minimum number of stars per snapshot $N_{thr}$.  We find the positive correlation between $f_{\mathrm{dim}}$ and $Q$ remains stable across the considered range.

The correlation dimension exhibits a stronger correlation with $Q$ than the box-counting dimension because it is more sensitive to the local clustering of stars and captures the scaling behavior of inter-point distances. In contrast, the box-counting dimension provides a more global measure of the overall spatial extent of the distribution, making it less responsive to local substructure. As a result, variations in $Q$, which primarily reflect changes in local clumpiness, are more directly reflected in the correlation dimension than in the box-counting dimension.

\section{Conclusions}
\label{Sec:Conclusions}

In this study, we investigated the evolution of fractality in centrally concentrated young star clusters and the role of massive stars in subcluster formation. Using a set of simulations of cluster formation from gas with different realizations of our initial conditions and numerical resolutions, we followed the gas collapse, star formation, and cluster assembly processes to analyze how these factors influence fractal structure and cluster concentration. We find the following conclusions.

\begin{itemize}
    \item Young clusters inherit fractal substructure from their parental molecular clouds, but in most cases this pattern is erased by dynamical relaxation at about $\sim2.5\,t_\mathrm{ff}$, regardless of the simulation resolution.
    \item Feedback from massive stars can trigger the formation of secondary subclusters, but its impact is not universal: whether feedback enhances fractality depends on both the stellar mass of stars producing significant feedback and the location and timing of their formation. 
    \item Interactions among subclusters, including their formation, merging, and dispersal, can sustain measurable fractality beyond $4\,t_\mathrm{ff}$.
    \item In the highest-resolution models (\texttt{n5s1} and \texttt{n6s1}), which evolve toward strongly centrally concentrated clusters ($Q\sim1.2$), the correlation dimension increases further in 3D while the box-counting dimension remains nearly constant, reflecting the higher sensitivity of the correlation dimension to dense stellar cores.
    \item The $Q$ parameter shows weak positive correlation with box-counting dimension, and moderate positive correlation with correlation dimension.
\end{itemize}

\begin{acknowledgements}
We thank the anonymous referee for the constructive comments and suggestions that helped improve the clarity and quality of this paper. The authors thank Sami Dib for sharing data on Galactic stellar clusters and Richard Parker for sharing N-body simulation results.
This research was funded by the Science Committee of the Ministry of Science and Higher Education of the Republic of Kazakhstan through grant AP26103591. AA acknowledges support from a Bolashaq International Scholarship. EA is partially supported by Nazarbayev University Faculty Development Competitive Research grant 040225FD4713. M-MML is partially supported by US National Science Foundation grant AST23-07950. XP acknowledges the financial support of the National Natural Science Foundation of China through grants 12573036 and 12233013 and the China Manned Space Program through grant  CMS-CSST-2025-A08.
\end{acknowledgements}

\bibliographystyle{bibtex/aa}
\bibliography{bibtex} 

\begin{appendix}
\onecolumn

\section{Simulations with no fractality}

\begin{figure*}[h!]
\centering
\includegraphics[width=0.75\textwidth]{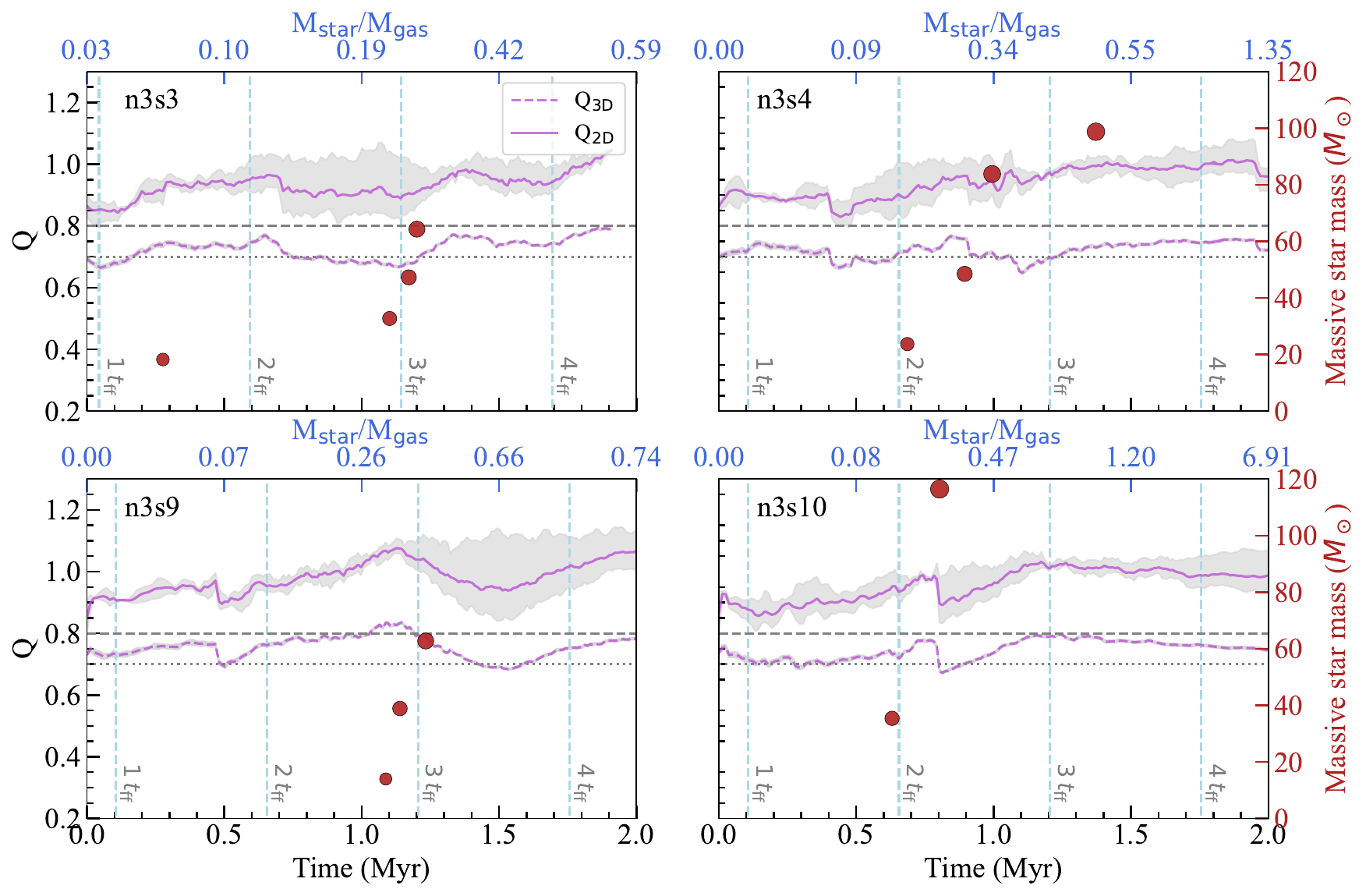}
    \caption{Evolution of the $Q$ parameter over time in models that initially show no fractality.
    All notations are the same as in Figure~\ref{fig:6Qs}.
   }
    \label{fig:4Qs}
\end{figure*}

In this appendix, we present simulations that do not develop fractal structure during their evolution. As shown in Figure~\ref{fig:4Qs}, the $Q$ parameter remains nearly constant throughout the simulation and stays in the regime associated with centrally concentrated stellar distributions. This indicates that the stellar distribution does not fragment into subclusters; instead, the systems evolve into centrally concentrated clusters while preserving a largely smooth global spatial distribution.

\begin{figure*}[h!]
\centering
\includegraphics[width=0.75\textwidth]{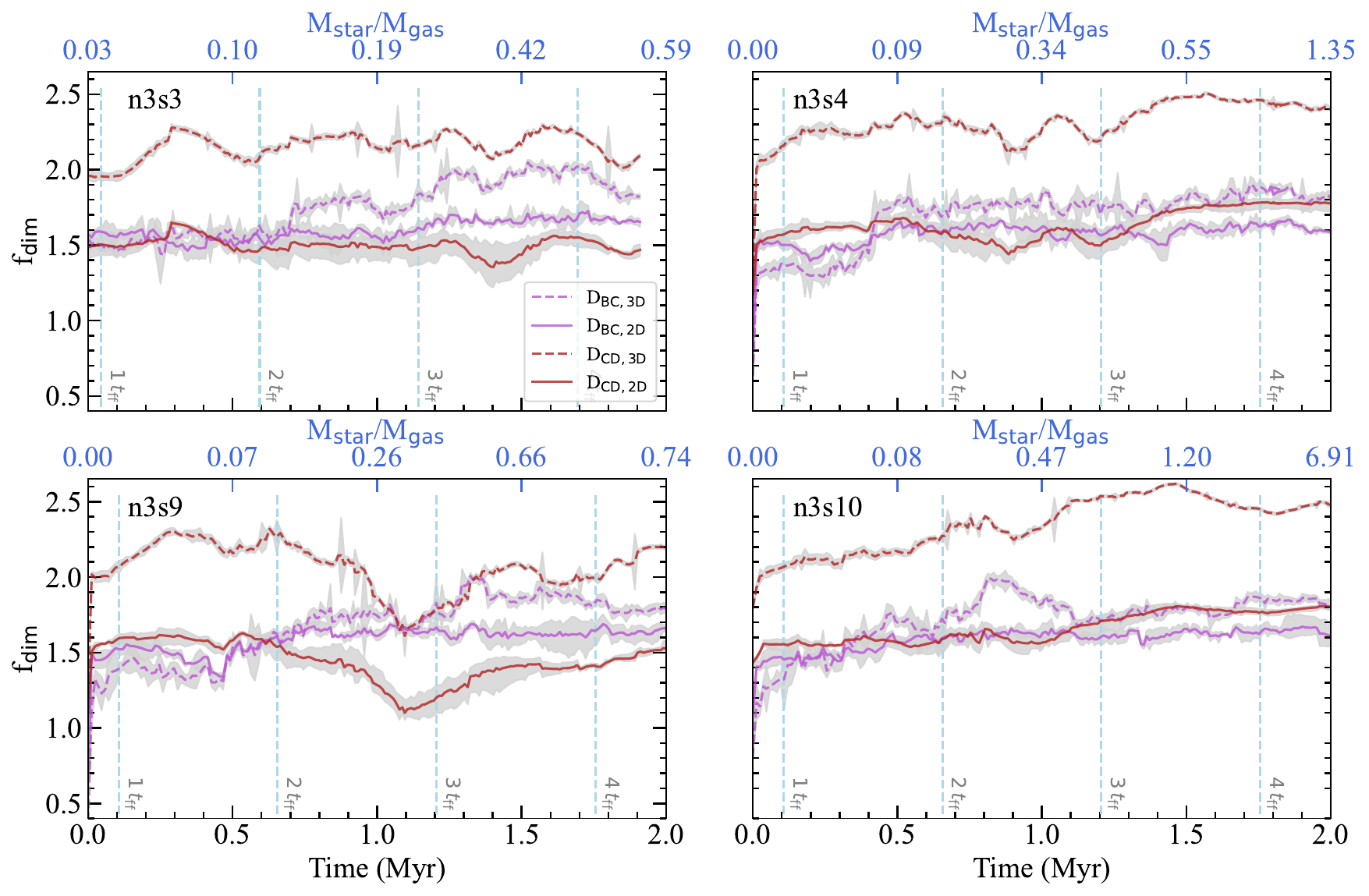}
    \caption{Evolution of the fractal dimension $f_{\mathrm{dim}}$ over time in models that initially show no fractality.  All notations are the same as in Figure~\ref{fig:6Ds}.  \label{fig:4Ds}}
\end{figure*}

The corresponding evolution of the fractal dimension $f_{\mathrm{dim}}$ is shown in Figure~\ref{fig:4Ds}. Overall, these models exhibit values of $f_{\mathrm{dim}}$ similar to those discussed in the main text. However, in contrast to simulations that develop fractal substructure, the fractal dimension in these models tends to plateau at slightly lower values.

\section{Uncertainty of $f_{\mathrm{dim}}$}
There are two primary sources of uncertainty in the estimation of the fractal dimension $f_{\mathrm{dim}}$: (i) uncertainties in the underlying data, and (ii) uncertainties inherent to the numerical estimation procedure itself. Data-related uncertainties typically arise in observational contexts due to measurement noise, instrumental limitations, or reconstruction effects, as discussed, for example, in~\citet{qin20253d}.

Since this work is based on simulation data, the first source of uncertainty is absent, leaving only methodological uncertainties associated with the estimation procedure.

In this study, the dominant methodological uncertainty arises from the choice of the fitting interval used to determine the slope of the log--log scaling relation. In particular, the inferred value of $f_{\mathrm{dim}}$ depends on the positions of the lower and upper bounds of the fitting window. To quantify the sensitivity of the fit to this choice, we systematically vary both bounds and compute the corresponding standard deviation of the fitted slope.

\begin{figure*}[h!]
\centering
\includegraphics[width=1\textwidth]{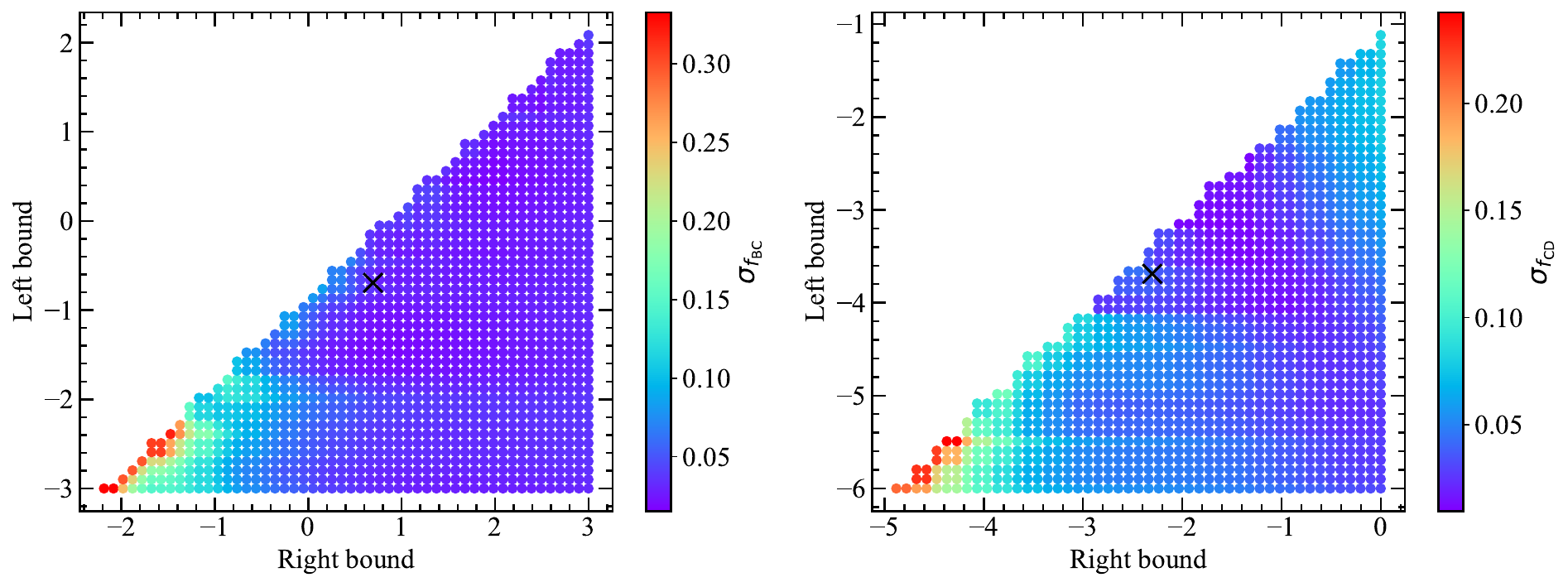}
\caption{Uncertainty map of the fitted fractal dimension for model \texttt{n6s1} at 1.2\,Myr. The colour scale shows the standard deviation of the fitted slope, $\sigma_f$, as a function of the left and right bounds of the fitting interval. The left panel corresponds to the box-counting estimate, while the right panel shows the correlation-dimension estimate. The black cross marks the fitting bounds adopted in the main analysis. The box-counting dimension is seen to be highly stable over a broad range of fitting intervals, whereas the correlation-dimension estimate shows a somewhat stronger, but still well-localized, dependence on the fitting bounds.}
\label{fig:fd_uncertainty}
\end{figure*}

The resulting uncertainty map is shown in Fig.~\ref{fig:fd_uncertainty}. The left panel corresponds to the box-counting estimate, while the right panel shows the correlation-dimension estimate. In both cases, the colour scale represents the standard deviation of the fitted slope, $\sigma_f$, evaluated over the explored fitting intervals.

We find that the box-counting estimate is particularly stable across a broad range of fitting bounds, as indicated by the extended low-$\sigma_{f_{\mathrm{BC}}}$ region in the left panel. This demonstrates that the box-counting dimension is only weakly sensitive to moderate shifts of the fitting interval and therefore provides a robust estimate of the underlying fractal structure.

The correlation-dimension estimate exhibits a somewhat stronger dependence on the choice of fitting bounds, although a clearly stable region is still present around the adopted fitting interval. This behaviour is expected, since the underlying log--log relation is not perfectly linear over all scales (see Fig.~1), and different fitting bounds probe slightly different local slopes.

In practice, the fitting interval used in the main analysis is selected by visually identifying the most linear portion of the scaling relation, i.e.\ the region that most closely follows a power-law behaviour. The crosses in Fig.~\ref{fig:fd_uncertainty} mark the adopted fitting bounds. These locations lie well within the low-uncertainty regions, supporting the robustness of the chosen fitting procedure.

We therefore conclude that, while the exact numerical values of $f_{\mathrm{dim}}$ retain some sensitivity to the fitting interval, the adopted estimates are robust and representative of the physically meaningful scaling regime.

\twocolumn

\end{appendix}

\end{document}